\documentclass[preprint, amsmath, amssymb, aps, pre]{revtex4-2}

\usepackage{graphicx}
\usepackage{dcolumn}
\usepackage{bm}
\usepackage{xcolor}
\usepackage{color}
\usepackage{caption}
\usepackage{subcaption}
\usepackage{amsmath}
\usepackage{amssymb}
\newcommand{\pnuc}{p_{\rm nuc}}
\newcommand{\phat}{\hat{p}_{\rm nuc}}
\newcommand{\tnuc}{t_{\rm nuc}}
\newcommand{\is}{Ising}
\newcommand{\mo}{model}
\newcommand{\nn}{nearest-neighbor}
\newcommand{\nuc}{nucleation}

\newcommand{\prob}{probability}
\newcommand{\lr}{logistic regression}
\newcommand{\lra}{long-range}
\newcommand{\ps}{pseudospinodal}

\newcommand{\ncop}{n_{\rm copies}}
\newcommand{\tint}{t_{\rm int}}
\newcommand{\nsuc}{n_{\rm suc}}
\newcommand{\cnn}{convolutional neural network}
\newcommand{\spin}{spinodal}
\newcommand{\dhh}{\Delta h}
\newcommand{\mf}{mean-field}

\begin{document}

\raggedbottom

\title{Predicting nucleation using machine learning in the Ising model}

\author{Shan Huang}
\email{sh2015@bu.edu}
\affiliation{Department of Physics, Boston University, Boston, Massachusetts 02215, USA}

\author{William Klein}
\email{klein@bu.edu}
\affiliation{Department of Physics, Boston University, Boston, Massachusetts 02215, USA}
\affiliation{Center for Computational Science, Boston
University, Boston, Massachusetts 02215, USA}

\author{Harvey Gould}
\affiliation{Department of Physics, Boston University, Boston, Massachusetts 02215, USA}
\affiliation{Department of Physics, Clark University, Worcester, Massachusetts 01610, USA}

\begin{abstract}
We use a \cnn\ (CNN) and two logistic regression models to predict the probability of nucleation in the two-dimensional Ising model. The three models successfully predict the probability for the \nn\ \is\ model for which classical nucleation is observed. The CNN outperforms the logistic regression models near the spinodal of the \lra\ Ising model, but the accuracy of its predictions decreases as the quenches approach the spinodal. Occlusion analysis suggests that this decrease is due to the vanishing difference between the density of the nucleating droplet and the background. Our results are consistent with the general conclusion that predictability decreases near a critical point.
\end{abstract}

\pacs{Valid PACS appear here}
\maketitle

\section{\label{sec:introduction}Introduction}

Many physical systems operate near or at a critical point, including earthquakes~\cite{earthquake1, earthquake2}, electrical activities in the brain~\cite{brain1, brain2, brain3}, the spread of forest fires~\cite{forestfire} and epidemics~\cite{disease}. It is well established that the predictability of local events in these systems is reduced due to the divergent spatial correlations at the critical point~\cite{earthquakepredict1, predict}. However, a recent study of the predictability of the size of events in the Olami-Feder-Christensen (OFC) model~\cite{OFCmodel} at criticality~\cite{OFCprediction} suggests that a divergent correlation length might not be the only reason for a decrease of predictability. It was found find that predictability in the OFC model decreases as criticality is approached and that prediction is possible only for large, nonscaling events~\cite{OFCprediction}. The main goal of this paper is to investigate in another context if the critical point imposes a limitation on the success of machine learning. We do so by predicting the probability of nucleation in the nearest-neighbor Ising model far from the critical point and near the spinodal, a critical point of the \mf\ Ising model.

Nucleation in the Ising model has been well studied. The nucleating droplets or clusters in classical nucleation are compact with a well defined boundary between the droplet and the background~\cite{classicalnuc}. Classical nucleation has been tested in the nearest-neighbor Ising model~\cite{testclassical}. Klein et al.~\cite{bigklein} have shown that nucleation near the spinodal critical point in the \lra\ Ising model is qualitatively different than classical nucleation. In contrast to classical nucleation, the nucleating clusters are ramified with a density equal to the background~\cite{spinodalnuc3}.

Machine learning has been very successful in various prediction problems. Hu et al.~\cite{application1} applied principal component analysis, an unsupervised learning clustering technique, to classify the disordered and ordered states and find the phase transition in the Ising and Blume-Capel models. Carrasquilla et al.~\cite{application2} used neural networks to make similar predictions for disordered systems that do not have order parameters. Pun et al.~\cite{OFCprediction} used a \cnn\ (CNN)~\citep{alexnet} to make predictions of the event sizes in the OFC model. It has been shown that a deep neural network can approximate arbitrarily complex functions~\cite{universalapproximation}, making it a powerful tool to study the predictability of various systems.

In this paper we apply several machine learning methods to predict the probability of nucleation for the \nn\ \is\ \mo\ and for the \lra\ \is\ \mo\ near the spinodal~\cite{footnote}. In Sec.~\ref{sec:data} we discuss the intervention method for generating data for the \nuc\ probability, which we use to train the machine learning models. In Sec.~\ref{sec:models} we introduce  two logistic regression models and a CNN model that we use to make predictions of the \nuc\ \prob. In Sec.~\ref{sec:result} we compare the performance of these models and show that CNN is more powerful and robust, but its predictability decreases as the system approaches the spinodal. In Sec.~\ref{sec:occlusion} we use occlusion analysis to identify the sensitive regions in the input Ising configurations and find that these regions overlap with the largest clusters found by a percolation mapping. These regions increase in size and becomes less defined as the system is quenched closer to the spinodal. We associate this observation with the emerging symmetry between the background fluctuations and the nucleation cluster near the spinodal. [xx could use a stronger sentence here xx]

\section{\label{sec:data}Data Acquisition}

Our simulations of the Ising model were performed using the Metropolis algorithm. We consider the \nn\ Ising model for which the critical temperature $T_c = 2/(\log(1~+~\sqrt{2}))$ and the long-range Ising model with interaction range $R \geq 10$ for which $T_c \approx 4$. The system is first equilibrated at temperature $T = 4/9 T_c$ with the magnetic field $h$ pointing up. We then flip the magnetic field to the down direction so that the system is metastable. The average lifetime of the metastable state is $\approx 10^4$ Monte Carlo (MC) steps per spin. Due to thermal fluctuations, clusters of down spins grow and vanish. Eventually, a cluster reaches the critical size at which there is a 50\% probability that the cluster will grow and fill the lattice.

Instead of identifying and monitoring the clusters during a run, which is computationally intensive, we estimate the nucleation time in our simulations as the time that the magnetization $m$ becomes smaller than a threshold value $m_0$. For the sizes of the system we consider, the actual nucleation time is only a few MC steps per spin before this time. For the long-range Ising model, we choose $m_0=0$. For the nearest-neighbor Ising model nucleation occurs much earlier than the time when $m$ first becomes negative. Instead, we determine the probability that $m_{t+1} < m_t$ and choose the value of $m_0$ to be the value of $m$ for which this probability is greater than 95\%. The values of $m_0$ for different values of the linear dimension $L$ are shown in Table~\ref{tab:irreversable}.

\begin{table}[t]
\begin{tabular}{|l|l|l|l|l|l|}
\hline
\multicolumn{1}{|c|}{$L$} & 10 & 30 & 50 & 70 & 100 \\ \hline
$m_0$ & 0.32 & 0.54 & 0.76 & 0.83 & 0.88 \\ \hline
\end{tabular}
\caption{The values of the threshold $m_0$ used to to estimate the nucleation time for the nearest-neighbor Ising model. The value of $m_0$ increases with $L$ because the radius of gyration of the critical droplet is a decreasing fraction of the system as $L$ increases.}
\label{tab:irreversable}
\end{table}

Our goal can be expressed as follows: given a two-dimensional configuration of spins as input, we wish to train a statistical model to predict the probability $\pnuc$ that the system will nucleate within the time $\tnuc$. We choose $\tnuc=50$ for $R=1$ and $\tnuc=25$ for $R \geq 10$. As discussed in Sec.~\ref{sec:intervention} we estimate $\pnuc$ by the intervention method, and and then use this data in Sec.~\ref{sec:dataset} to train the various statistical models to predict $\pnuc$.

\subsection{Intervention method}
\label{sec:intervention}

\begin{figure}[t]
\includegraphics[width = 0.8\textwidth]{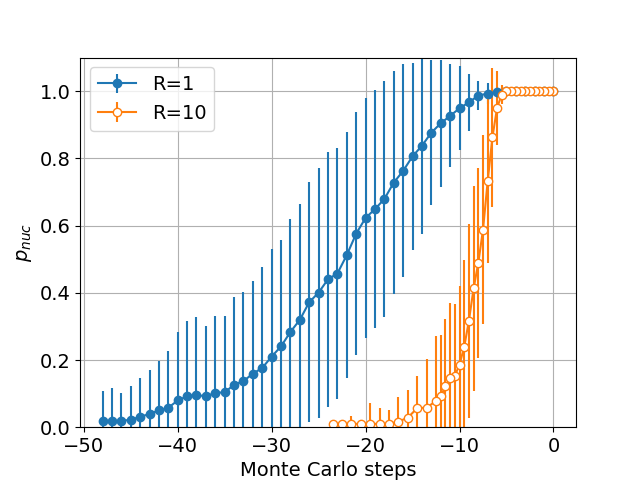}
\caption{\label{fig:intervention}The time dependence of the nucleation probability $\pnuc$. The time $t=0$ corresponds to the time at which $m$ first satisfies $m \leq m_0$. Nucleation occurred before $t=0$. Results are shown for $R=1$, $\Delta h=0.70$, $L=10$ (\textcolor{blue}{$\bullet$}) and $R=10$, $\Delta h=0.055$, $L=100$ (\color{orange}{$\circ$}).}
\end{figure}

It is convenient to take $t=0$ to be the time that $m \leq m_0$ and the simulation is stopped. We save the spin configurations at every MC time step for the previous $\tnuc$ MC updates per spin. The intervention method proceeds by making $\ncop=100$ copies of the spins at the intervention time $\tint < 0$. We then run each copy with a different random number seed and determine the number of copies $\nsuc$ that reach $m \leq m_0$ within the time $t_{\mathrm{nuc}}$. The nucleation probability $\pnuc$ at time $t$ is taken to be $\pnuc(t) = \nsuc(t)/\ncop$. Typical results for $\pnuc$ from the intervention method is given in Fig.~\ref{fig:intervention}.

We did not monitor if the copies nucleated at the same position and time as the original run. Because the size of the system in our simulation is relatively small, the probability of two nucleating droplets in the system at roughly the same time is small.

\subsection{Dataset preparation}
\label{sec:dataset}

There is much more data for which $\pnuc\approx 0$ or $\pnuc\approx 1$ in our original dataset and than for intermediate values of $\pnuc$. To ensure that the predictions do not favor the most common values of $\pnuc$, we discard data according to how often their corresponding values of $\pnuc$ occur, so that there is a a uniform distribution of $\pnuc$ in the training dataset.

\begin{figure}[t]
\includegraphics[width=0.95\textwidth]{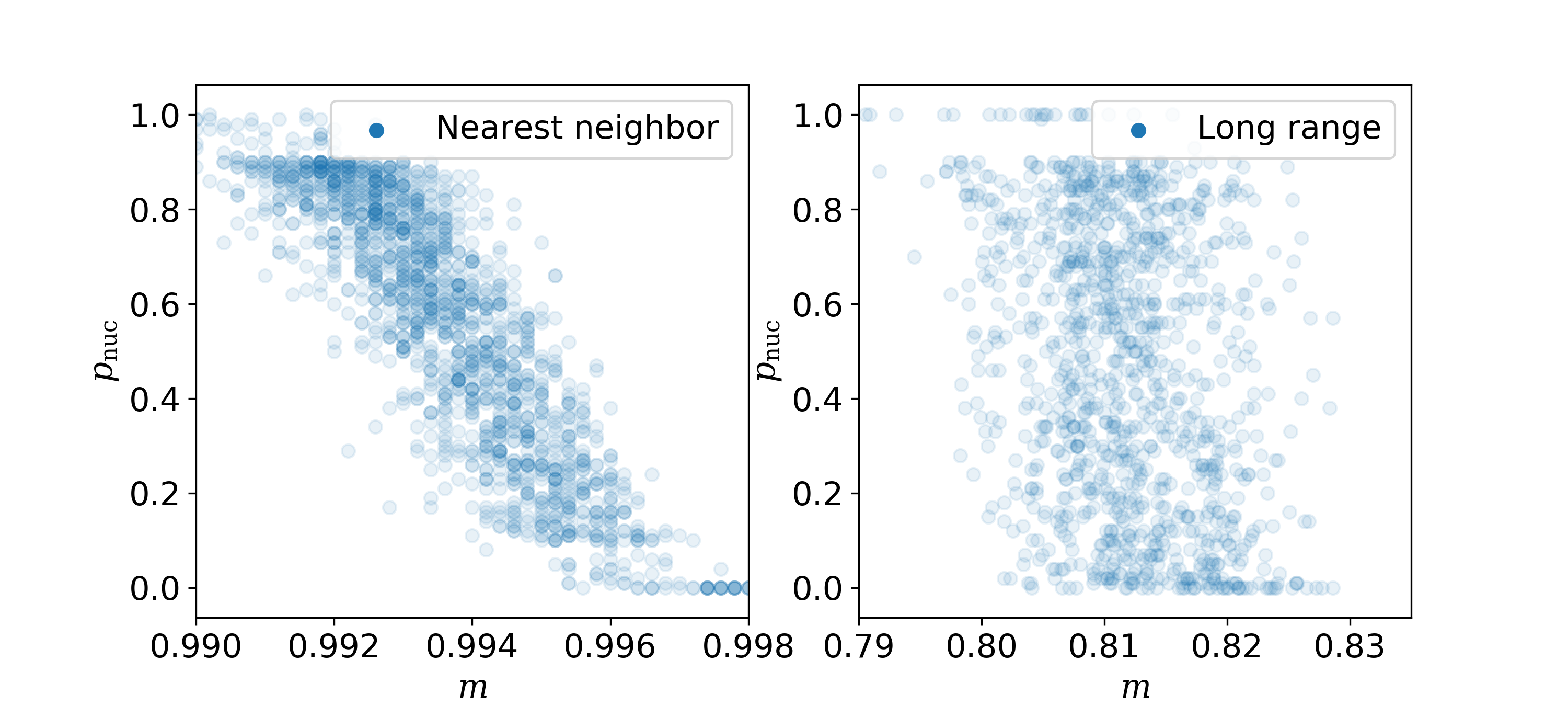}
\caption{The nucleation probability $\pnuc$ determined by the intervention method versus the magnetization $m$ for (a) $R=1$, $h = 0.50$, $L=100$ and (b) for $R=10$, $h=1.21$, $L=100$. Recall that $m>0$ in the metastable state. For classical nucleation, $m$ is a reasonable indicator of the probability of nucleation. For $R=10$ there is no obvious correlation between $m$ and $\pnuc$.}
\label{fig:p_nucvsm}
\end{figure}

Before we apply machine learning methods, we first need to determine if there is an obvious relation between the nucleation probability and the magnetization $m$. Figure~\ref{fig:p_nucvsm} shows that for $R=1$, the value of $m$ is a reasonable indicator of $\pnuc$. [xx so what? what do you do with this info? why mention it here? xx] For $R=10$, there is little correlation between $m$ and $\pnuc$. The squared Pearson correlation coefficient is $r_1^2 = 0.37$ and $r_{10}^2=0.14$, respectively.

We divide the dataset into training and testing datasets. The former is used to train the machine learning models and the latter is used to test the model's performance by comparing the predicted probability $\phat$ with the value $\pnuc$ obtained by the intervention method.

Because the machine learning methods we use do not take periodic boundary conditions into account, we translate each spin configuration so that the center of mass of the largest cluster is at the center of the simulation box. The clusters are determined using the Coniglio-Klein percolation mapping~\cite{percolationmapping}.

\section{\label{sec:models}Machine learning models}

We applied three models: spin-based logistic regression, features-based logistic regression, and \cnn\ (CNN).

\subsection{\label{sec:spin-logreg}Spin-based logistic regression}

The logistic regression models can be summarized as
\begin{equation}
\phat(X) = \frac{1}{1+\exp(-\sum^N_{i=1}{W_i X_i})}.
\label{eq:logreg}
\end{equation}
where $X$ is the input and $W$ is the weight vector. During training, the logistic regression models adjusts the weight vector $W$ so that $\phat$ is as close to $\pnuc$ as possible. For the spin-based \lr\ method, the spin configuration is represented by a one-dimensional $N$-element array $X$, with each element corresponding to a spin. Each element $W_i$ can interpreted as the importance of spin $i$ in the input. This naive model works well for image recognition tasks when the features are easy to capture and the input size is not very big.

\subsection{Features-based logistic regression}

In the spin-based logistic regression model each spin is considered to be an independent feature. This assumption is questionable because nucleation is a collective phenomenon. The idea of features-based logistic regression is to construct a higher level representation of a configuration with some a priori chosen geometric quantities. We use this representation as the input in Eq.~\eqref{eq:logreg} with $N$ replaced by the number of features. In this way the model is more concise, but less information is retained about the system.

We choose the magnetization $m$, the average distance to the center of mass of the largest cluster $R^{(1)}$ and the higher moments $R^{(k)}$ as our features. The higher moments are given by
\begin{align}
R^{(k)} & = \Big (\frac{1}{N_s}\sum^{N_s}_i r_i^k\Big)^{1/k},
\label{eq:Features} \\
\noalign{\noindent with}
X & = (m, R^{(1)}, R^{(2)}, \ldots R^{(k_{\max})}),
\end{align}
where $N_s$ is the number of spins in the stable (down) direction, and $r_i$ is the distance of (down) spin $i$ to the center of mass of the largest cluster. We include $R^{(k)}$ up to $k_{\max}=9$. The model's representation is the same as Eq.~\eqref{eq:logreg}, with $X$ now represented by the geometric features in Eq.~\eqref{eq:Features}.

\subsection{Convolutional Neural Network}

A \cnn\ conducts convolution operations in a small local region called a ``filter,'' and sweeps this filter through the entire input. This behavior allows it to detect local geometrical correlations of the spin configurations with translational invariance, meaning that the model gives exactly the same response, no matter where the features of interest appears in the input~\cite{footnote2}. We used a multi-layer CNN. Between each layer of a convolutional operation, we insert a maxpool layer to extract the important information. For $R=1$ we use a 2-layer CNN, with eight and sixten $5 \times 5$ filters respectively. For $R=10$ we use a three layer CNN with 16, 32, 64 filters respectively. The maxpool layer takes a maximum of every $2\times2$ window. To prevent over-fitting, we place a dropout layer after each CNN layer, with a dropout rate of 0.15~\cite{dropout}.

The models are trained with ADAM optimizer~\cite{adam}, with learning rate 0.0003, $\beta_1 = 0.99$, and $\beta_2 = 0.999$ to minimize the mean square error between $\pnuc$ and $\phat$ defined as
\begin{equation}
\mbox{MSE} = \langle [\pnuc-\phat]^2 \rangle.
\label{eq:mse}
\end{equation}

\section{\label{sec:result}Results}

\subsection{Predicting the probability of classical nucleation}

\begin{figure}[t]
\includegraphics[width=0.6\textwidth]{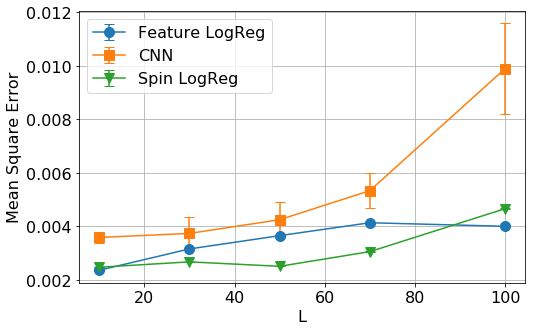}
\caption{\label{fig:NNPerformance} Dependence of the mean square error of spin-based logistic regression ({\color{green}{$\blacktriangledown$}}), features-based logistic regression (\textcolor{blue}{$\bullet$}), and CNN ({\color{orange}{$\blacksquare$}}) on the linear dimension $L$ of the \nn\ \is\ model. The performance of the two logistic regression methods depends only weakly on $L$, but the performance of the CNN decreases with increasing $L$.}
\end{figure}

We applied the three models to the nearest-neighbor Ising model with $h=0.50$ and values of $L$ between 10 to 100 to determine the dependence of the performance of each model on $L$. Each model was averaged over a 10-fold cross-validation~\cite{crossvalidation}. From Fig.~\ref{fig:NNPerformance} we see that both logistic regression models show only a weak dependence on $L$, but the performance of the CNN significantly decreases with increasing $L$.

Because the largest cluster is at the center of the lattice and its size is independent of $L$, increasing $L$ does not yield more useful information. For features-based logistic regression, the surrounding background makes no contribution to the calculation of the geometric features. For example, for spin-based logistic regression, the model learns to reduce the weights of the background spins in Eq.~\eqref{eq:logreg} because they do not contribute to the model's predicted values of $\phat$. In contrast, because the filter for the CNN sweeps through the entire input configuration, these values overwhelm the CNN with useless information, which we suspect is the main reason for the decrease in CNN's performance with increasing $L$. That is, the CNN tunes the weight on the filters, which sweep through the entire grid.

\subsection{Predictions for the long-range \is\ model}
\label{sec: CNNCloseToSpinodal}

\begin{figure}[h]
\includegraphics[width=11cm]{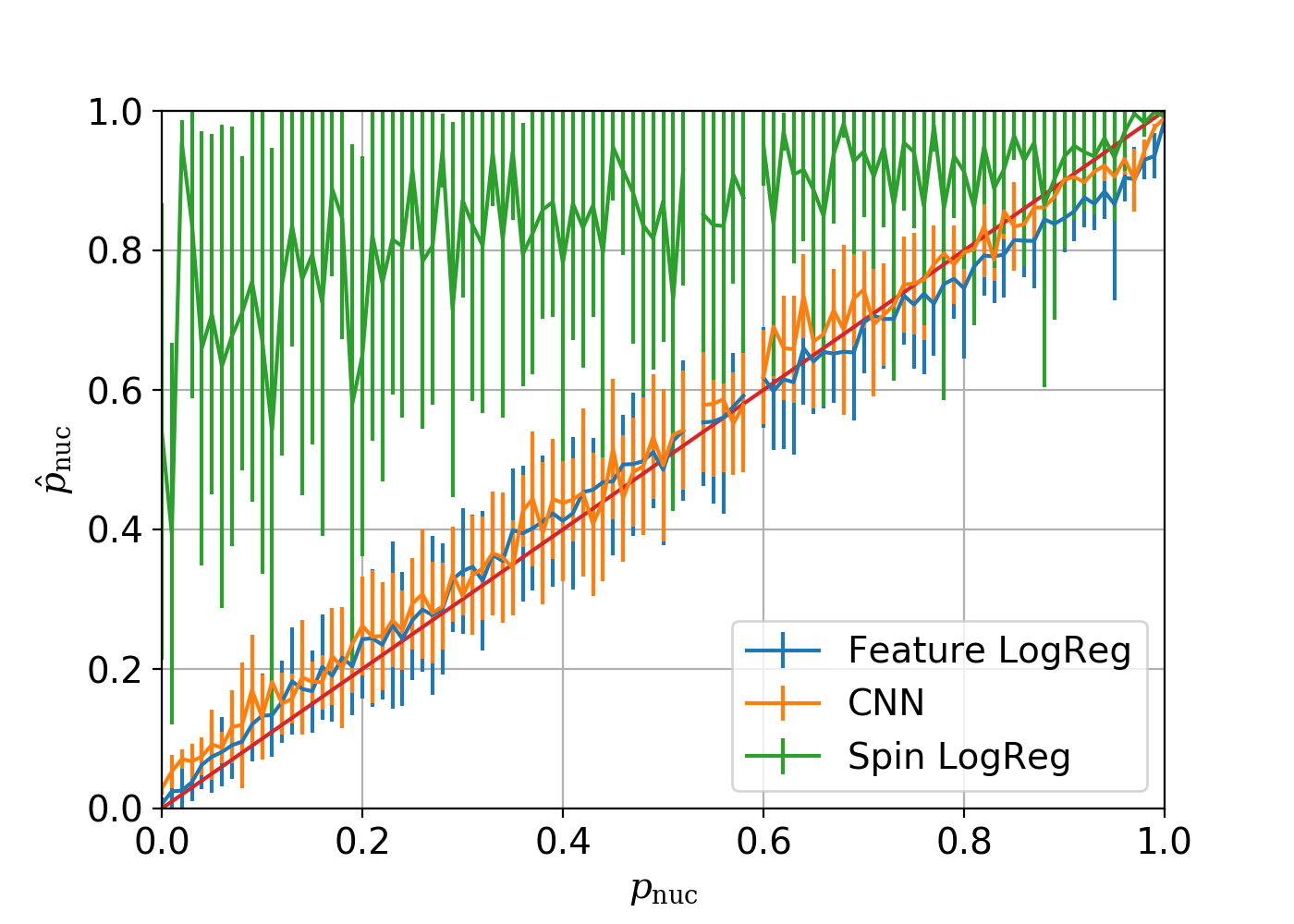}
\caption{Comparison of $\pnuc$ and the values of $\phat$ generated by the three machine learning models near the \spin\ with $\Delta h = 0.06$, $R=10$, and $L=100$. The diagonal line represents perfect prediction. The predictions of features-based logistic regression and CNN are reasonably accurate. The mean-square-error of CNN is equal to 0.0046 compared to the mean-square-error of 0.0084 for features-based logistic regression.}
\label{fig:ConfusionPlot}
\end{figure}

Figure~\ref{fig:ConfusionPlot} compares the values of $\phat$ predicted by the three models to the actual nucleation probabilities $\pnuc$ for $R=10$ and $\Delta h = 0.06$. We see that CNN demonstrates the best performance, with the mean-square-error of CNN and features-based logistic regression equal to 0.0046 and 0.0084, respectively. Both models perform much better than spin-based logistic regression.

As mentioned in Sec.~\ref{sec:introduction}, the closer the OFC model is to the noise-induced critical point, the less predictable is the size of the events given the stress configuration of the system~\cite{OFCprediction}. We now determine if a similar loss of predictability occurs as the spinodal in the \lra\ Ising model is approached.

Because the probability of nucleation and thus the lifetime of the metastable state depends on the Ginzburg parameter $G \sim R^2 \Delta h$ (in two dimensions)~\cite{bigklein}, we increase $R$ as we decrease $\Delta h$ to keep $G$ fixed at $G=6$. In this way the lifetime of the metastable state is at least $10^4$\,MC steps per spin as we increase $h$ toward $h_s = 1.27$, the value of the spinodal critical point for $T=4T_c/9$.

\begin{figure}[h]
\includegraphics[width=0.6\textwidth]{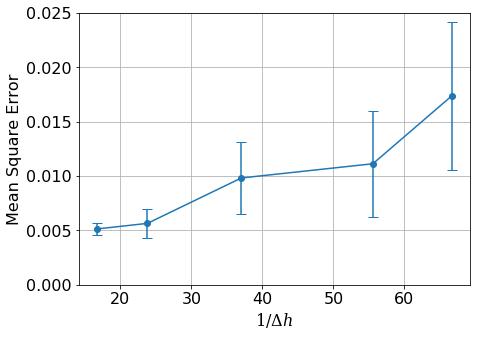}
\caption{The performance of the CNN as the magnetic field $h$ approaches $h_s$ for fixed linear dimension $L=200$. The interaction range $R$ is chosen so that the Ginzburg parameter $G$ is constant with $G=6$. As $h$ approaches $h_s$, the performance of the CNN model becomes less robust.}
\label{fig:CNNCTS}
\end{figure}

We set $L=200$ to ensure that the input size is not a factor when we compare the performance of the same CNN model for different values of $R$. Each CNN model was trained on 7000 data points and validated on 1000 data points. Our results for the mean square error are shown in Fig.~\ref{fig:CNNCTS}. The error is calculated based on a 10-fold cross-validation. We see that as $h$ approaches $h_s$, the CNN model exhibits worse performance. This decrease is discussed in more detail in Sec.~\ref{sec:occlusion}.

\subsection{\label{sec:ST}Temporal and spatial input}

\begin{figure}[h]
\includegraphics[width=0.8\textwidth]{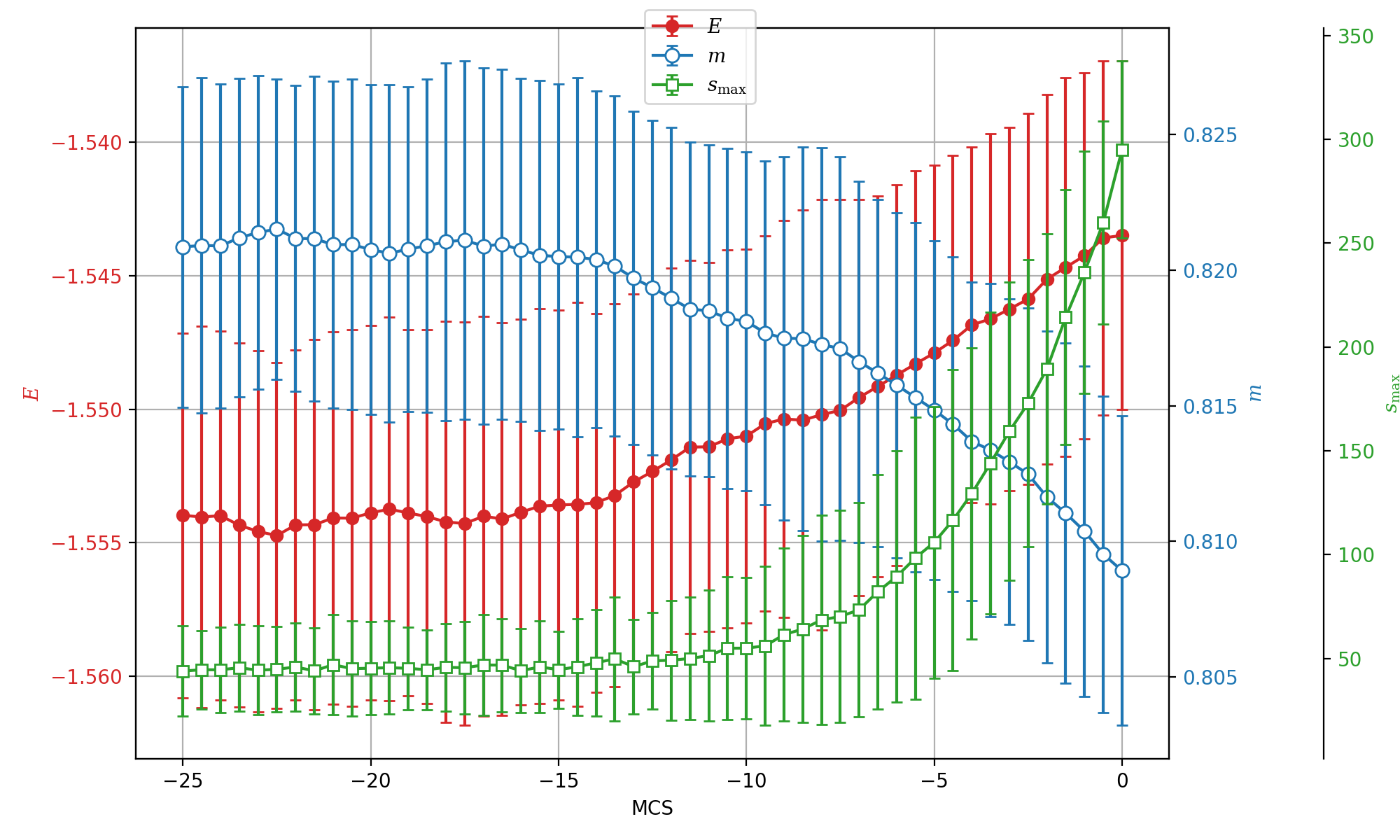}
(\caption{\label{fig:me}The evolution of the energy per spin $E/N$(\textcolor{red}{$\bullet$}), the magnetization $m$ ({\color{blue}{$\circ$}}), and the number of spins in the largest cluster $s_{\max}$ ({\color{green}{$\Box$}}) in the Ising model with $R=10$, $L=200$, $\Delta h = 0.06$.
The evolution is averaged over 1000 nucleation events.}
\end{figure}

So far we have used spatial information to generate predictions for $\pnuc$. In the following we discuss our results using temporal information only and a combination of temporal and spatial information. Figure~\ref{fig:me} shows the evolution of the magnetization $m$ and the energy per spin $E/N$ before nucleation. The large fluctuations of $m$ and $E/N$ make it difficult to use their evolution to make predictions. However, if we use the Coniglio-Klein percolation mapping to determine the largest cluster $s_{\max}$, we see that the evolution of the number of spins in the largest cluster has a much smaller variance

\begin{figure}[t]
\includegraphics[width=0.8\textwidth]{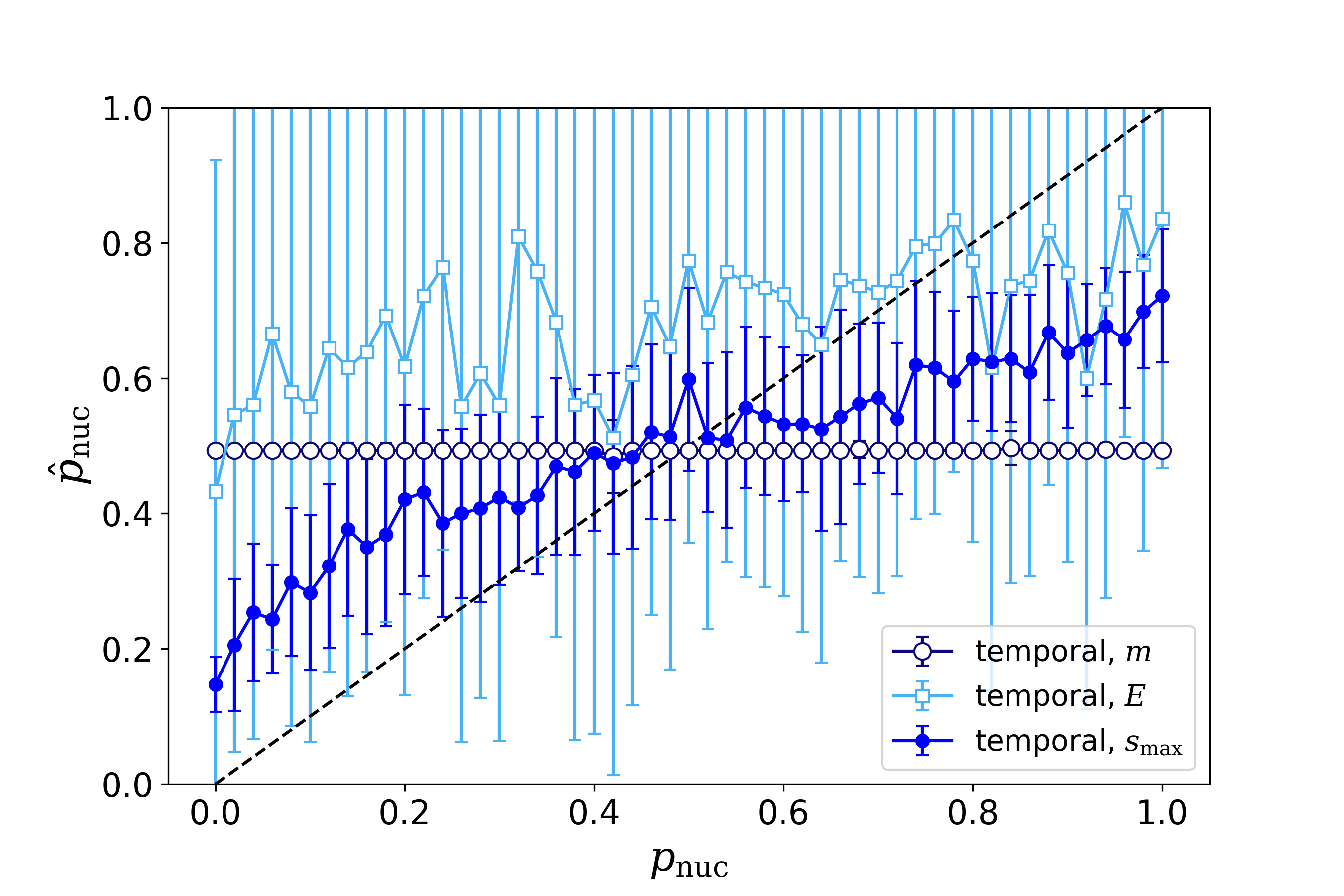}
\caption{\label{fig:temporal}Comparison of using $m$ ({\color{blue}{$\circ$}}) $E/N$ ({\color{blue}{$\Box$}}), or $s_{\max}$ ({\color{blue}{$\bullet$}}) as input for determining $\phat$. The predicted results from the CNN using $s_{\max}$ shows some positive correlation with the true value ($\mbox{MSE} = 3.96 \times 10^{-2}$). The evolution of $m$ and $E/N$ does not lead to accurate predictions ($\mbox{MSE} = 9.05 \times 10^{-2}$ and $9.40 \times 10^{-2}$ respectively)
}
\end{figure}

We chose $m$, $E$, or $s_{\max}$ as input to train three one-dimensional CNN models to predict $\phat$ near the spinodal with $\Delta h = 0.06$, $R=10$, and $L=200$. The input corresponds to the last 25\,MC steps per spin prior to $t=0$, similar to how we trained the CNN using only spatial information. The performance of the three models is compared in Fig.~\ref{fig:temporal}. As expected, the evolution of the directly measurable quantities $m$ and $E$ is not helpful for determining $\phat$. The predicted results using $s_{\max}$ show some positive correlation with the true value, but not as strong as when spatial information is used. We performed the same task closer to the spinodal and found that the mean square error increases as the \spin\ is approached, as it did using only spatial information, although using spatial information gave much better predictability.

\begin{figure}[t]
\includegraphics[width=0.8\textwidth]{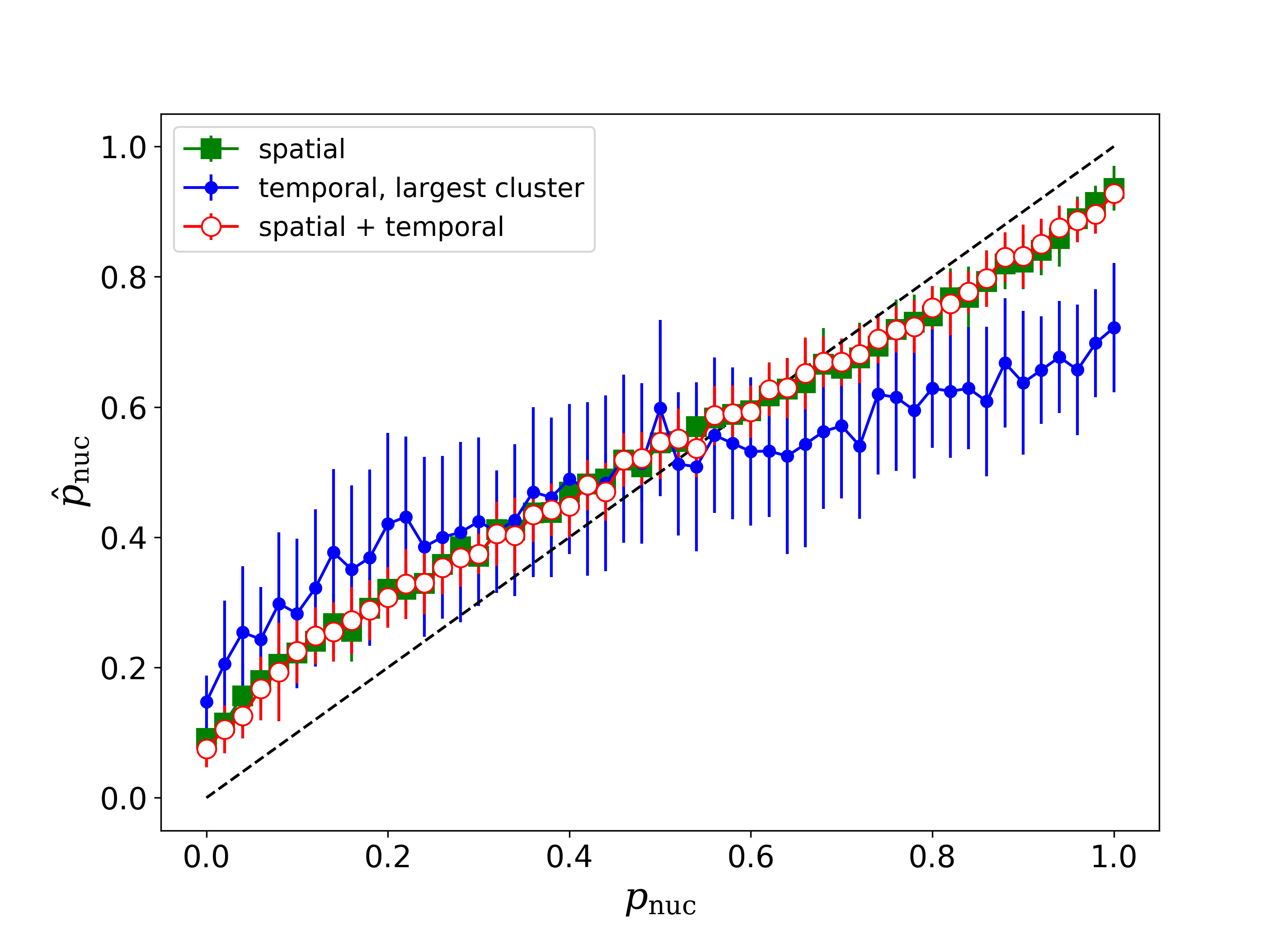}
\caption{\label{fig:ST}Comparison of the performance of pure spatial ({\color{green}{$\blacksquare$}}, $\mbox{MSE} = 4.43 \times 10^{-3}$), pure temporal ({\color{blue}{$\bullet$}}, $\mbox{MSE} = 3.96 \times 10^{-2}$), and spatial + temporal ({\color{red}{$\circ$}}, $\mbox{MSE} = 6.12 \times 10^{-3}$) CNN models. Combining spatial and temporal information made the predictions somewhat less accurate than using spatial information only.}
\end{figure}

We also constructed a machine learning model that combines spatial and temporal information with the hope of decreasing the model's mean square error. The spatial module takes the spin configuration at the time of interest and encodes it with a deep two-dimensional CNN. The temporal module takes the history of the past 25\,MC steps per spin at the time of interest, and encodes it with a one-dimensional CNN. The two outputs are then concatenated and go through two fully connected linear layers to produce $\phat$. As shown in Fig.~\ref{fig:ST}, the combination of spatial and temporal information does not increase the accuracy of the model's predictions. Because the temporal information is a weak indicator of nucleation, including it in the input adds uncertainty, which confuses the model.

\section{\label{sec:occlusion}Important features and occlusion analysis}


\begin{figure}[t]
\includegraphics[width=0.5\textwidth]{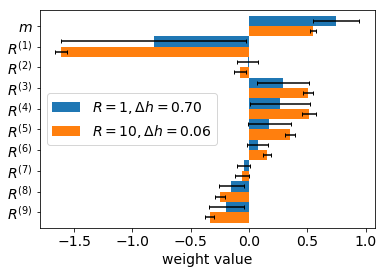}
\caption{The weights corresponding to each geometric feature in features-based logistic regression (averaged using cross-validation). The most important features are the magnetization $m$ and the first moment $R^{(1)}$.}
\label{fig:GeoFeatWeights}
\end{figure}

As discussed in Sec.~\ref{sec:spin-logreg}, the weight vector $W$ in the logistic regression models represents the importance of each feature in the input. As shown in Fig.~\ref{fig:GeoFeatWeights}, the most important features of the features-based \lr\ model are the magnetization $m$ (positively correlated) and the first moment $R^{(1)}$ (negatively correlated). These correlations are expected: the more spins in the stable (down) direction, the more likely that nucleation will occur. Also the smaller the density of the down spins in the vicinity of the largest cluster, the less likely nucleation will occur.

The values of the higher moments $R^{(k)}$ are much more important for $R=10$ than for $R=1$. The greater importance of the higher moments is consistent with the prediction of spinodal nucleation theory that the nucleating cluster has a much larger spatial extent.

Occlusion analysis is commonly used to visualize and understand what the \cnn\ has learned~\cite{occlusion}. For the Ising model this analysis is done by choosing a small region of the input and replacing the spins in the occluded region by zeros, and observing how much the predictions of the model vary when fed the original and occluded inputs. The bigger the difference, the more important the region. This process is repeated many times to obtain an occlusion sensitivity map~\cite{occlusion}, which illustrates the importance of each part of the input.

\begin{figure}[h]
\includegraphics[width=0.4\textwidth]{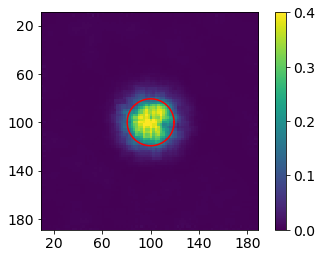}
\caption{Occlusion map for $R=10$, $\Delta h = 0.06$, and $L=200$. The red circle is the average radius of gyration of the biggest cluster, which overlaps with the high sensitivity region of the occlusion map.}
\label{fig:OccExample}
\end{figure}

Figure~\ref{fig:OccExample} shows the occlusion sensitivity map for $R=10$, $\Delta h =0.06$, $L=200$ averaged over the validation dataset. As expected, the most important region in the lattice is at the center, because the input is centered about the largest cluster. Note that the radius of the sensitive region is comparable to the value of $\overline{R}_g=19.4$, shown by the red circle. This consistency indicates that the machine acquires useful geometric information from the region occupied by the largest cluster.

\begin{figure}[h]
\begin{subfigure}{0.49\textwidth}
\includegraphics[width=7.5cm]{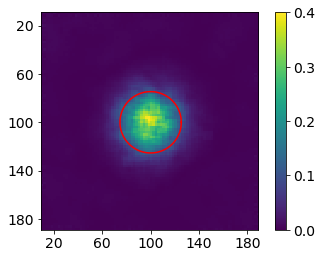}\\
\caption{[$\Delta h=0.042$, $R=12$, $\overline{R}_g= 27$. }
\label{fig: R=12}
\end{subfigure}
\begin{subfigure}{0.49\textwidth}
\includegraphics[width=7.5cm]{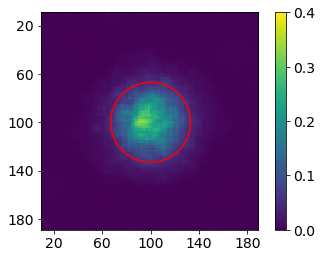}\\
\caption{$\Delta h=0.027$, $R=15$, $\overline{R}_g= 37$.}
\label{fig: R=15}
\end{subfigure}
\begin{subfigure}{0.49\textwidth}
\includegraphics[width=7.5cm]{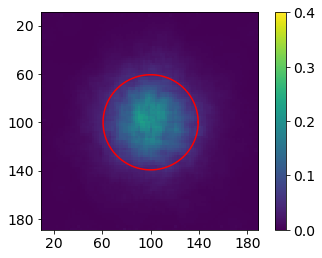}\\
\caption{$\Delta h=0.018$, $R=18$, $\overline{R}_g = 48$.}
\label{fig: R=18}
\end{subfigure}
\begin{subfigure}{0.49\textwidth}
\includegraphics[width=7.5cm]{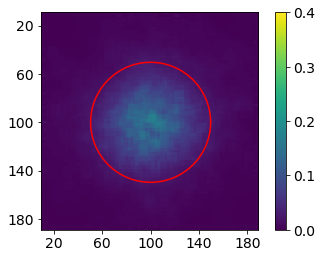}\\
\caption{$\Delta h=0.015$, $R=20$, $\overline{R}_g= 57$.}
\label{fig: R=20}
\end{subfigure}
\caption{\label{fig:occresult}Occlusion sensitivity maps for the Ising model with $R=10$ and $L=200$ for decreasing values of $\dhh$, keeping $G$ fixed at $G=6$. The radius of the circular region enclosed by the red line is equal to $R_g$, the mean radius of gyration of the largest cluster. Note that the size of the occlusion sensitive region becomes larger but less intense, similar to the increasing size and decreasing density of the largest cluster.}
\end{figure}

Occlusion analyses were also done on systems closer to the \spin\ for $10 < R \leq 20$ [xx not $\leq$? xx] and $L=200$ (see Fig.~\ref{fig:occresult}). As $\dhh \to 0$, the spatial extent of the sensitive region increases and the sensitivity decreases. The size of the sensitive region is comparable to the radius of gyration $R_g$ of the largest cluster in the system.

\begin{figure*}
\begin{subfigure}{1.0\columnwidth}
\includegraphics[width=9cm]{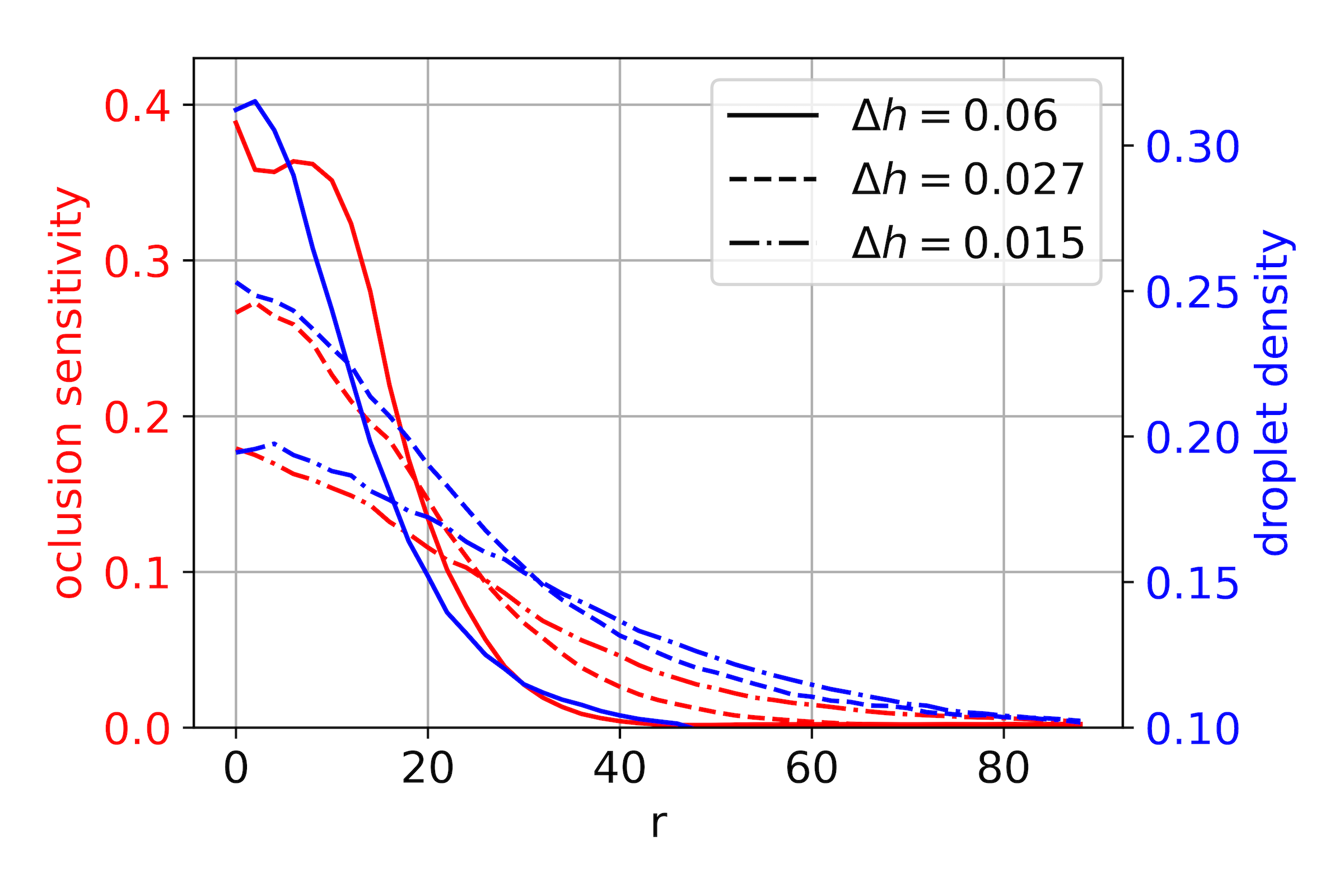}
\end{subfigure}
\begin{subfigure}{1.0\columnwidth}
\includegraphics[width=9cm]{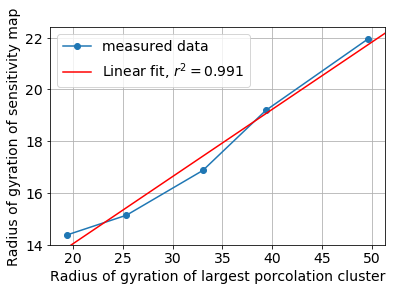}
\end{subfigure}
\caption{(a) Dependence of the occlusion sensitivity (red lines) and the density of the nucleating cluster (blue dashed lines) on $r$, the distance from the center of mass of the largest cluster. (b) The radius of gyration of the occlusion sensitivity region is strongly correlated with the radius of gyration of the largest cluster.}
\label{fig:OccVsDensity}
\end{figure*}

Near the spinodal the nucleating droplet consists of a dense core and a ramified halo~\cite{spinodalnuc3}. As $\dhh \to 0$, the core shrinks, the halo grows, and the density of the cluster approaches the density of the background (see Fig.~\ref{fig:OccVsDensity})~\cite{dropletdensity}. The greater spatial extent of the nucleating droplet is consistent with the larger radius of the occlusion sensitive region. The decrease of the density difference between the nucleating droplet and the background also explains the decrease in sensitivity of the occlusion analysis and	 the decreasing predictability of the CNN.

\begin{figure}
\includegraphics[width=0.6\textwidth]{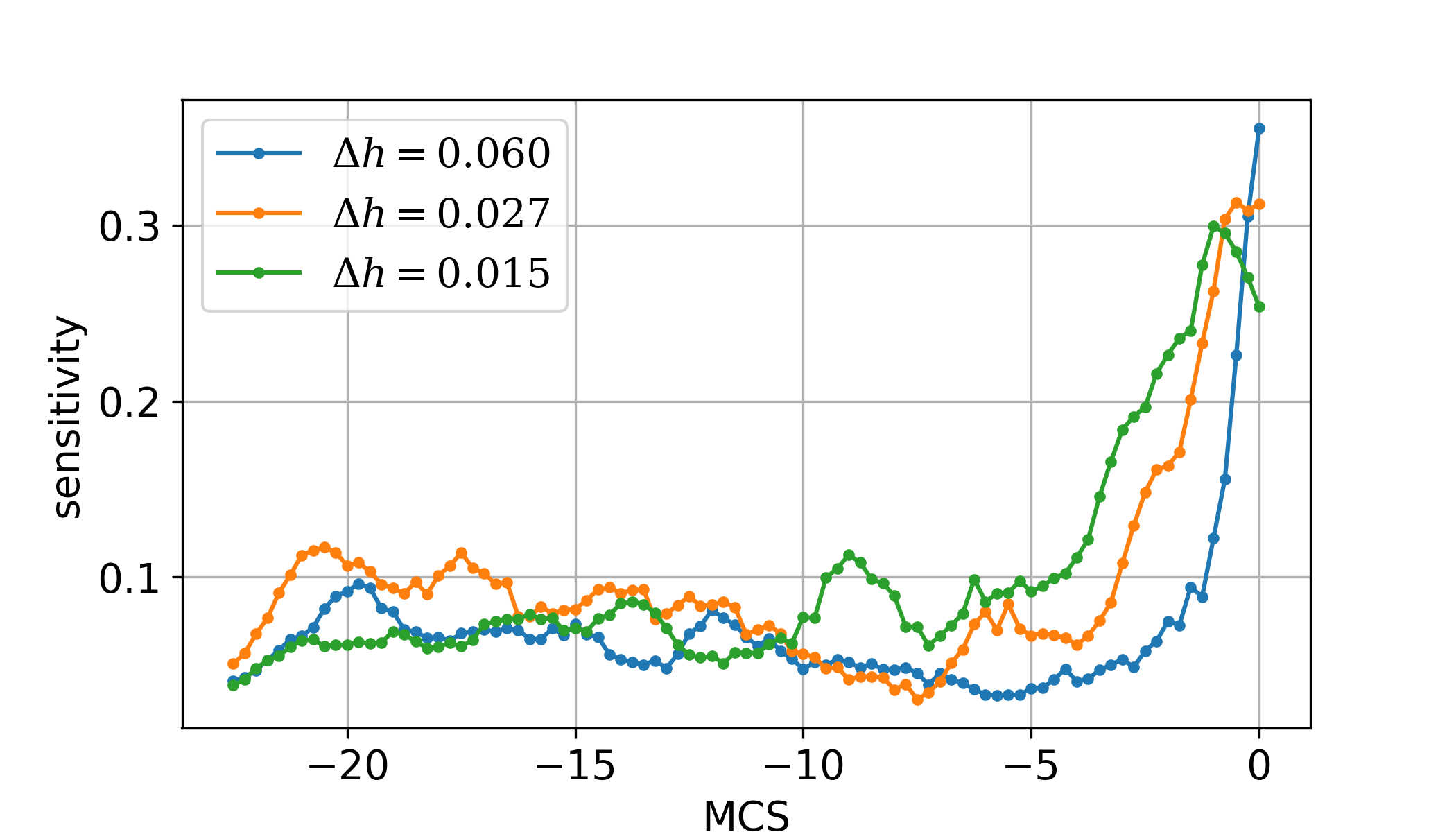}
\caption{\label{fig:STOcclusion}Occlusion sensitivity map for the history of the size of the largest cluster as input. The sensitive region extends further into the past as $h \rightarrow h_s$.}
\end{figure}

We did a similar occlusion analysis on the temporal CNN described in Sec.~\ref{sec:ST}. We observe that the sensitive region in time extends further into the past, which is consistent with the increasing importance of critical slowing down as the system approaches the spinodal (see Fig.~\ref{fig:STOcclusion}).

\section{Summary}

We applied several machine learning methods to predict the probability of both classical and spinodal nucleation in the Ising model. We found that the CNN gives accurate predictions for classical nucleation and for nucleation close, but not too close, to the spinodal. We related the decrease of sensitivity found by our occlusion analysis and the decreasing difference between the density of the nucleating droplet and the background to the decrease in predictability as the \spin\ is approached. Our results indicate that systems near critical points may be inherently unpredictable by statistical models. Nevertheless, the CNN is still a reasonable predictor of nucleation for the \nn\ Ising model and for for the \lra\ Ising model if the spinodal is not approached too closely. And applications of the CNN can be  generalized to other physical systems  to make  predictions of nucleation in systems such as dense liquids.

\begin{acknowledgements}

We would like to give special thanks to Chon-Kit Pun, for his previous work on predicting event size in OFC model using CNN and Sakib Matin, Thomas Tanzin and Timothy Khouw for their helpful comments.

\end{acknowledgements}


\begin{thebibliography}{99}

\bibitem{earthquake1} P. Bak and C. Tang, ``Earthquakes as a self-organized critical phenomenon,'' J. Geophysical Research: Solid Earth {\bf 94}, 15635 (1989).

\bibitem{earthquake2}I.~G. Main and F.~H. Al-Kindy, ``Entropy, energy, and proximity to criticality in global earthquake populations,'' Geophysical Research Lett. {\bf 29}, 25 (2002).

\bibitem{brain1}J.~M. Beggs and D. Plenz, ``Neuronal avalanches in neocortical circuits,'' J. Neuroscience {\bf 23}, 11167 (2003).

\bibitem{brain2}D.~R. Chialvo, ``Emergent complex neural dynamics,'' Nature Physics \textbf {6}, 744 (2010).

\bibitem{brain3}J. Hesse and T. Gross, ``Self-organized criticality as a fundamental property of neural systems,'' Frontiers in Systems Neuroscience \textbf{8}, 166 (2014).

\bibitem{forestfire}W. Song, F. Weicheng, W. Binghong, and Z. Jianjun, ``Self-organized criticality of forest fire in China,'' Ecological Modelling \textbf {145}, 61 (2001).

\bibitem{disease}H. Saba, J. Miranda, and M. Moret, ``Self-organized critical phenomenon as a $q$-exponential decay—avalanche epidemiology of dengue,'' Physica A \textbf {413}, 205 (2014).

\bibitem{earthquakepredict1}I. Main, ``Earthquake prediction: concluding remarks, Nature {1} (1999).

\bibitem{predict}D. Sornette, {\sl Critical Phenomena in Natural Sciences}, 2nd ed. (Springer, 2006).

\bibitem{OFCmodel}K. Christensen and Z. Olami, ``Variation of the Gutenberg-Richter $b$ values and nontrivial temporal correlations in a spring-block model for earthquakes,'' J. Geophysical Research: Solid Earth \textbf {97}, 8729 (1992).

\bibitem{OFCprediction}C.-K. Pun, S. Matin, W. Klein, and H. Gould, ``Prediction in a driven-dissipative system displaying a continuous phase transition,'' Phys. Rev. E {\bf 101}, 022102 (2020).

\bibitem{bigklein}W. Klein, H. Gould, N. Gulbahce, J. Rundle, and K. Tiampo, ``Structure of fluctuations near mean-field critical points and spinodals and its implication for physical processes, Phys. Rev. E \textbf {75}, 031114 (2007).

\bibitem{classicalnuc}R. Becker and W. D\"oring, ``Kinetische behandlung der keimbildung in \"ubers\"attigten d\"ampfen, Annalen der Physik \textbf{416}, 719 (1935).

\bibitem{testclassical}V. A. Shneidman, K. A. Jackson, and K. M. Beatty, ``On the applicability of the classical nucleation theory in an Ising system,'' J. Chem. Phys. {\bf 111}, 6932 (1999).

\bibitem{spinodalnuc3}C. Unger and W. Klein, ``Initial-growth modes of nucleation droplets,'' Phys. Rev. B \textbf {31}, 6127 (1985).


\bibitem{application1}W. Hu, R.~R. Singh, and R.~T. Scalettar, ``Discovering phases, phase transitions, and crossovers through unsupervised machine learning: A critical examination,'' Phys. Rev. E \textbf {95}, 062122 (2017).

\bibitem{application2}J. Carrasquilla and R.~G. Melko, ``Machine learning phases of matter,'' Nature Physics \textbf {13}, 431 (2017).

\bibitem{alexnet}A. Krizhevsky, I. Sutskever, and G.~E. Hinton, ``Imagenet classification with deep convolutional neural networks,'' in \emph {Advances in Neural Information Processing Systems} (2012), pp. 1097--1105. 

\bibitem{universalapproximation}B.~C. Cs\'aji, ``Approximation with artificial neural networks,'' Faculty of Sciences, E\"otv\"os Lor\'and University \textbf {24}, 48 (2001).

\bibitem{footnote}The spinodal is well defined only in the mean-field limit. For finite interaction ranges $R\gg 1$, the spinodal is replaced by a \ps. For simplicity, we will refer to the \ps\ as a spinodal. See N. Gulbahce, H. Gould, and W. Klein, ``Zeros of the partition function and pseudospinodals in long-range Ising models,'' Phys. Rev. E{\bf 69}, 036119 ??(2004).

\bibitem{percolationmapping}A. Coniglio and W. Klein, ``Clusters and Ising critical droplets: a renormalisation group approach, J. Phys. A: Mathematical and General \textbf {13}, 2775 (1980).

\bibitem{footnote2}In this case it is not necessary to position the largest cluster at the center. However, when we do the occlusion analysis in Sec.~\ref{sec:occlusion}, it is convenient to place the largest cluster at the center so that we can average over many configurations.

\bibitem{dropout} N. Srivastava, G. Hinton, A. Krizhevsky, I. Sutskever, and R. Salakhutdinov, ``Dropout: a simple way to prevent neural networks from overfitting,'' J. Machine Learning Research \textbf {15}, 1929 (2014).

\bibitem{adam}D.~P. Kingma and J. Ba, ``Adam: A method for stochastic optimization,'' Third International Conference on Learning Representations (2015), arXiv 1412.6980 (2014).

\bibitem{crossvalidation}R. Kohavi et al., ``A study of cross-validation and bootstrap for accuracy estimation and model selection,'' in Ijcai Vol.~{\bf 14}, Montreal, Canada (1995), pp. 1137--1145.

\bibitem{occlusion} M.~D. Zeiler and R. Fergus, ``Visualizing and understanding convolutional networks,'' in {\emph European Conference on Computer Vision} (Springer, 2014), pp. 818--833.

\bibitem{dropletdensity}K. Binder, ``The Monte Carlo method for the study of phase transitions: A review of some recent progress,'' J. Computational Phys. \textbf {59}, 1 (1985).

\bibitem{rg}P. Mehta, M. Bukov, C.-H. Wang, A. G. R. Day, C. Richardson, C. K. Fisher, and D. J. Schwab, ``A high-bias, low-variance introduction to Machine Learning for physicists,'' Phys. Reports {\bf 810}, 1 (2019).

\end{thebibliography}
\end{document}